\documentclass[sn-nature]{sn-jnl}
\usepackage{graphicx}%
\usepackage{multirow}%
\usepackage{amsmath,amssymb,amsfonts}%
\usepackage{amsthm}%
\usepackage{mathrsfs}%
\usepackage[title]{appendix}%
\usepackage{xcolor}%
\usepackage{textcomp}%
\usepackage{manyfoot}%
\usepackage{booktabs}%
\usepackage{algorithm}%
\usepackage{algorithmicx}%
\usepackage{algpseudocode}%
\usepackage{listings}%
\usepackage{url}
\usepackage{subcaption}
\usepackage{svg}
\usepackage{nicefrac, xfrac}
\newgeometry{left=2cm,right=2cm, top=2cm, bottom=2cm}

\theoremstyle{thmstyleone}%

\theoremstyle{thmstyletwo}%

\theoremstyle{thmstylethree}%

\begin{document}

\title[Article Title]{Accurate machine learning force fields via experimental and simulation data fusion}

\author[1]{\fnm{Sebastien} \sur{Röcken}}
\author*[1,2]{\fnm{Julija} \sur{Zavadlav }}\email{julija.zavadlav@tum.de}

\affil[1]{Multiscale Modeling of Fluid Materials, Department of Engineering Physics and Computation, TUM School of Engineering and Design, Technical University of Munich, Germany}
\affil[2]{Munich Data Science Institute, Technical University of Munich, Germany}

\abstract{Machine Learning (ML)-based force fields are attracting ever-increasing interest due to their capacity to span spatiotemporal scales of classical interatomic potentials at quantum-level accuracy. They can be trained based on high-fidelity simulations or experiments, the former being the common case. However, both approaches are impaired by scarce and erroneous data resulting in models that either do not agree with well-known experimental observations or are under-constrained and only reproduce some properties. Here we leverage both Density Functional Theory (DFT) calculations and experimentally measured mechanical properties and lattice parameters to train an ML potential of titanium. We demonstrate that the fused data learning strategy can concurrently satisfy all target objectives, thus resulting in a molecular model of higher accuracy compared to the models trained with a single data source. The inaccuracies of DFT functionals at target experimental properties were corrected, while the investigated off-target properties remained largely unperturbed. Our approach is applicable to any material and can serve as a general strategy to obtain highly accurate ML potentials.}

\maketitle

\section*{INTRODUCTION}\label{sec1}
With their ability to accelerate the discovery of new materials and decipher the properties of existing materials, Molecular Dynamics (MD) simulations have become a cornerstone of material science~\cite{pilania2022recent}. Nevertheless, the true capability is often hindered by the accuracy vs. efficiency trade-off of traditional approaches. Ab initio MD provides high-accuracy predictions at low computational efficiency, while the contrary holds for the MD simulations based on classical force fields. In theory, Machine Learning (ML) approaches~\cite{hart2021machine, Vlachas2021accelerated} and, in particular, ML potentials~\cite{Friederich2021Machine,mueller2020machine, Mishin2021Machine,zuo2020performance} can overcome this compromise due to the multi-body construction of the potential energy with unspecified functional form. In practice, the success of ML potentials hinges primarily on the training data, the source of which can be either simulations or experiments, or both. 

Typically the former source is used with ab-initio calculations providing energy, forces, and potentially virial stress (target labels) for different atomic configurations (inputs)~\cite{behler2007generalized,erhard2022a,gasteiger2020fast,schutt2017schnet,unke2019PhysNet,batzner20223,musaelian2023learning}. Such a setup, also known as bottom-up learning, has the benefit of straightforward training and should result in ML potentials that reproduce all properties of the underlying model. However, generating ab-initio training data that is sufficiently accurate, large, and broad (without distribution shift) is challenging.

CCSD(T) (coupled cluster with single, double, and perturbative triple excitations) method, regarded as the gold-standard of electronic structure theory, is generally computationally infeasible for large dataset generation. Thus, most ML potentials are trained on the more affordable but less accurate Density Functional Theory (DFT) calculations. These are not always in quantitative agreement with experimental predictions, and consequently, neither are ML potentials trained on DFT data. For example, a recent ML-based model of titanium~\cite{wen2021specialising} does not quantitatively reproduce the experimental temperature-dependent lattice parameters and elastic constants. For these properties, it achieved a similar level of agreement with experiments as the classical MEAM (modified embedded atom method) potential~\cite{Lee2001second}. Deviations in the phase diagram predictions are also frequent~\cite{dickel2020neural, bartok2018machine, rosenbrock2021machine,li2020complex}. In all cases, these deviations were attributed to DFT inaccuracies. To approach the CCSD(T) level accuracy, transfer learning~\cite{smith2019approaching} or $\Delta$-learning~\cite{ramakrishnan2015big} techniques, exploiting a large DFT and a small CCSD(T) dataset, can be used. 

Nevertheless, DFT training data is, albeit cheaper, still computationally expensive, and an optimal selection of atomic configurations is needed for diverse and non-redundant training data. Typically, training datasets are carefully prepared and contain specialized sub-datasets based on the target application, such as surfaces, defects, lattice distortions, thermal displacements, configurations along the phase transformation pathways, etc.~\cite{botu2017machine,huan2017a,takahashi2017conceptual,zong2018developing} Alternatively, an active learning approach~\cite{smith2018less,kostiuchenko2019impact,podryabinkin2017active,zhang2019active,Sivaraman2020machine} is used, where the dataset is increased on the fly during training. These methods require robust uncertainty quantification scheme, which remains problematic for Neural Network (NN)-based potentials~\cite{Tan2023single, Thaler2023scalable,kahle2021quality,Zhu2022fast,musil2019fast,imbalzano2021uncertainty}.

Apart from the dataset size, the system size (number of atoms per configuration) can also play a significant role in the optimal model components and hyperparameters and, consequently, the resulting trained model~\cite{gasteiger2022gemnet-oc}. Due to the cubic scaling of DFT implementations, the average number of atoms is typically below one hundred for dense systems under periodic boundary conditions. It is questionable whether long-range interactions~\cite{anstine2023machine} can be learned from such databases, considering the recent finding that features related to interatomic distances as large as 15~\AA\; can play an essential role in describing non-local interactions~\cite{kabylda2023efficient}.

The difficulties of ab-initio data generation can be circumvented if ML potentials are instead trained top-down, i.e., on experimental data~\cite{thaler2021learning,wang2023learning,Navarro2023Top,frohlking2020toward}. While experimental data is also scarce, potentially laborious to obtain, and contains measurement errors, the obtained information per data sample is much larger compared to bottom-up learning. Experimentally observable properties of a system are in simulations computed as an ensemble average, i.e., averaged over a very large number of atomic configurations. This fact also complicates training since it requires running forward simulations to calculate the properties and, in principle, subsequent gradient backpropagation through the simulation. Automatic differentiation~\cite{baydin2018automatic} and recent end-to-end differentiable software~\cite{schoenholz2020jax, doerr2021torchmd,wang2022dmff} have made such endeavors technically possible. In practice, backpropagation through the simulation is unfeasible for properties that require long simulations due to issues such as memory overflow, exploding gradients, and high computational costs~\cite{thaler2021learning,ingraham2019learning,wang2020differentiable}. However, for time-independent properties, these issues can be avoided with Differentiable Trajectory Reweighting (DiffTRe) method~\cite{thaler2021learning} that, rather than backpropagating through the trajectory, employs a reweighting technique. For a test case diamond system, the method yielded an ML potential that reproduced the target experimental mechanical properties at ambient conditions. Yet, for out-of-target phonon density of states, substantially different results were obtained for different random initializations, showcasing that the high-capacity ML potentials are under-constrained when trained on a handful of experimental observations~\cite{thaler2021learning}. Combining both simulation and experimental data sources, an idea used for decades to construct classical force fields~\cite{purja2009development} and recently also a two-body correction to a fixed ML potential~\cite{Sakib2023machine}, should, therefore, yield the best approach also for ML potentials. 

In this work, we demonstrate, for the first time, how ML potentials can be trained simultaneously on simulation and experimental data. In particular, we train a Graph Neural Network (GNN) potential for titanium on DFT calculated energies, forces, and virial stress for various atomic configurations and experimental mechanical properties and lattice parameters in the temperature range of 4 to 973 K. We then test the resulting model that faithfully reproduces all target properties on several out-of-target properties, i.e., phonon spectra, liquid phase structural and dynamical properties. Surprisingly, out-of-target properties are only slightly affected by the combined training approach, revealing a remarkably large capacity of the state-of-the-art ML potentials.

%%%%%%%%%%%%%%%%%%%%%%%%%%%%%%%%%%%%%%%%%%%%%%%%%%%%%%%%
\section*{RESULTS AND DISCUSSION}\label{sec2}
\subsection*{Fused data training approach}
\begin{figure}[!htb]
\includegraphics[trim={0cm 3.5cm 0 0},clip,width=1.\linewidth]{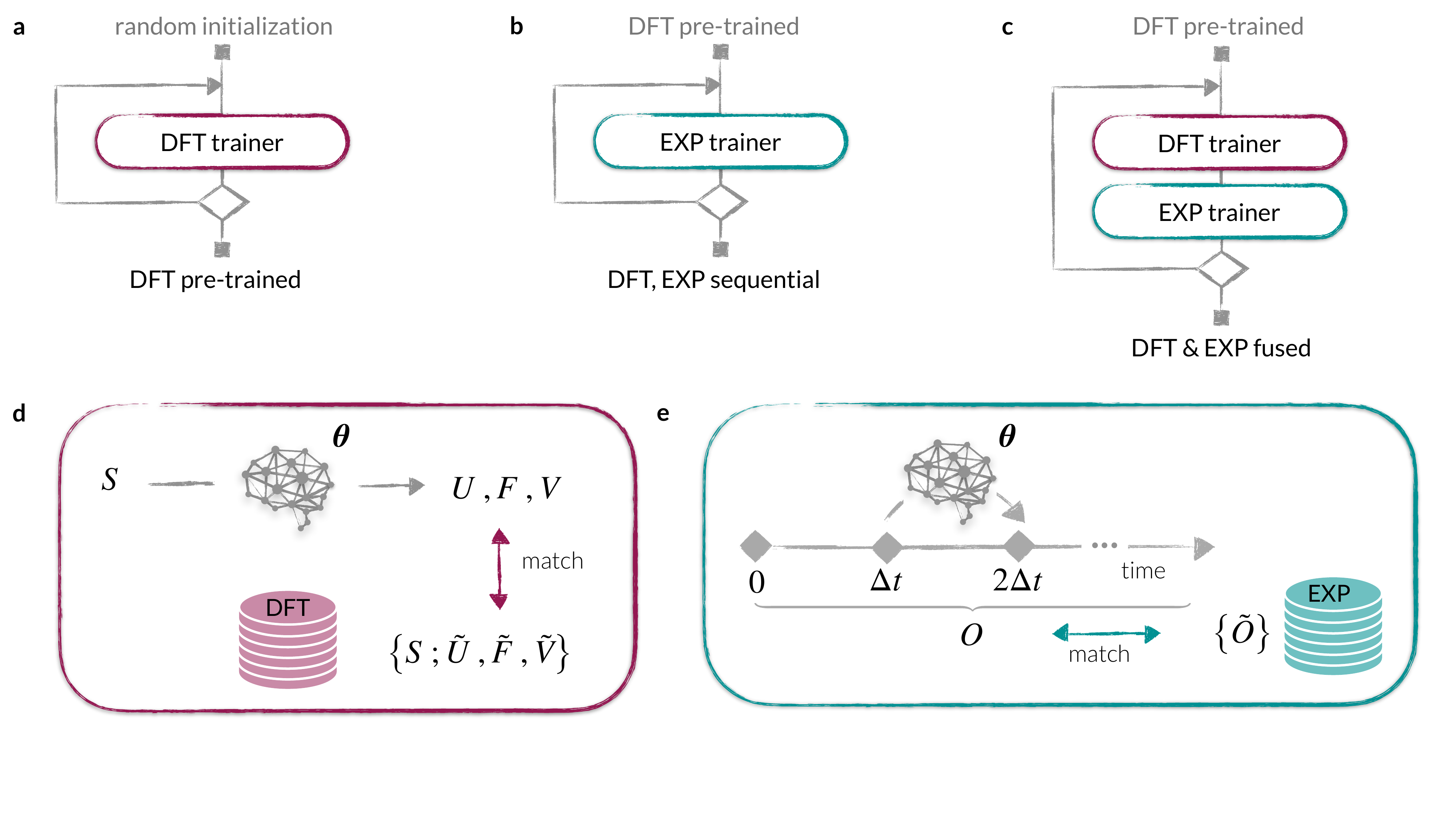}  
\caption{Investigated models. The DFT pre-trained model (\textbf{a}) is trained only with DFT trainer (\textbf{d}), which optimizes the parameters of the ML potential to match the reference DFT potential energy $\tilde{U}$, forces $\tilde{F}$, and virial $\tilde{V}$ for different atomic environments $S$. For the DFT, EXP sequential model (\textbf{b}), the ML potential is initialized with the parameters of the DFT pre-trained model and trained with EXP trainer (\textbf{e}), where the ML potential is trained to reproduce experimental observables $\tilde{O}$. EXP trainer requires simulations since the observables are not a direct output of the ML model but computed as a time average over the simulated trajectory. The DFT \& EXP fused model (\textbf{c}) is obtained by alternating between the DFT and EXP trainer, starting from the DFT pre-trained model.}
\label{fig:Scheme}
\end{figure}

A concurrent training on the DFT and experimental data can be achieved by iteratively employing both a DFT trainer and an EXP trainer (Figure~\ref{fig:Scheme}). 
%%%%%%%%%%%%%%
%%%%%%%%%%%%%%
The former involves a standard regression problem. The ML potential takes as an input atomic configuration $S$ and predicts the potential energy $U$ from which the forces on all atoms $F$ and virial stress tensor $V$ are computed by differentiating with respect to atoms' positions. The parameters $\theta$ are modified using batch optimization for one epoch to match the ML potential's predictions and the target values in the DFT database. We reuse the previously published DFT calculations for titanium~\cite{wen2021specialising,TiDATA}. The DFT database consists of 5704 samples. It includes equilibrated, strained, and randomly perturbed hcp, bcc, and fcc titanium structures, as well as configurations obtained via high-temperature MD simulations and an active learning approach. Further details are in the Supplementary Information.

The EXP trainer, on the other hand, performs optimization of parameters $\theta$ for one epoch such that the properties of titanium (observables) computed from the ML-driven simulation's trajectory match experimental values where the gradients are computed with the DiffTRe method~\cite{thaler2021learning}. We consider temperature-dependent, solid-state elastic constants of hcp titanium as target experimental properties. Elastic constants of titanium were measured experimentally at 22 different temperatures in the range of 4-973~K~\cite{simmons1971single}. Nevertheless, we select only the following four temperatures: 23, 323, 623, and 923~K for the experimental training database. With this choice, we reduce the computational cost per epoch and include our expectation that the models will be, to some degree, temperature transferable. The elastic constants are evaluated in the NVT ensemble, where the box size is set according to the experimentally determined lattice constants~\cite{souvatzis2007anomalous} (see Supplementary Information). Thus, by adding the additional target of zero pressure, we indirectly match also the experimental lattice constants. 

To investigate the impact of DFT and EXP trainers, we compare three different approaches; (i) the DFT pre-trained model, employing only the DFT trainer (ii) the DFT, EXP sequential model, employing only the EXP trainer, and (iii) the DFT \& EXP fused model, obtained with the alternating use of the DFT and EXP trainers. For the last two approaches, the parameters of the ML potential are not initialized randomly but with the values of the DFT pre-trained model. This allows us to circumvent the use of prior potentials, typical for top-down learning~\cite{thaler2021learning,wang2020differentiable}. The prior potentials are simple classical potentials added to the ML potential to avoid unphysical trajectories and, therefore, slow learning in the initial learning stage. 

%%%%%%%%%%%%%%%%%%%%%%%%%%%%%%%%%%%%%%%%%%%%%%%%%%%%%%%%
\subsection*{Simultaneously learning DFT and experimental target properties}
We compute the energy, force, and virial errors on the DFT test dataset (Table~\ref{tab:RMSE_InitAndBulk}) for all three investigated models.
%%%%%%%%%%%%%%
\begin{table}[h]
\begin{tabular}{@{}llll@{}}
\toprule
 & \multicolumn{3}{c}{RMSE/MAE} \\
Approach  & Energy [meV/atom] & Force  [meV/Å]  &  Virial [meV/atom] \\
\midrule
DFT pre-trained          & 6.0/4.4     & 92.5/62.1    & 406.6/261.2  \\
DFT, EXP sequential    & 385.1/384.9   & 123.6/83.8     & 401.5/267.1 \\
DFT \& EXP fused    & 7.9/6.2      & 111.2/76.6     & 405.6/263.2 \\
\bottomrule
\end{tabular}
\caption{Root Mean Square Error (RMSE) and Mean Absolute Error (MAE) of energy, force, and virial predictions computed on the test DFT dataset.}\label{tab:RMSE_InitAndBulk}
\end{table}
%%%%%%%%%%%%%%
For the DFT pre-trained model, the obtained energy error is below 43~meV, generally accepted within the chemistry community as the chemical accuracy~\cite{Faber2017Prediction}. In Supplementary Table S3, we additionally show the errors for a portion of the test dataset containing only strained and perturbed hcp or bcc samples. The force and virial errors are an order of magnitude lower when high-temperature configurations are excluded, demonstrating the greater difficulty of fitting the out-of-equilibrium structures. We compare favorably with the previously published ML-based potential model~\cite{wen2021specialising} for the force errors, while the energy errors are somewhat higher. However, precedence can be given to energy, force, or virial error by changing the weights of the loss function (Eq.~\ref{eq:DFTloss}). We give a higher emphasis on the forces as these are relevant for carrying out MD simulations. 

When training on both DFT and experimental data (DFT \& EXP fused model), the errors are only slightly increased compared to training only on DFT data (DFT pre-trained model). An increase is expected as the model has to satisfy both DFT and experimental objectives, which are partially conflicting due to the DFT inaccuracies as well as experimental errors. The fact that the errors do not change drastically indicates that the DFT errors in energy, force, and virial predictions are minor. Nevertheless, a small difference in force prediction can amount to large differences in MD simulations and subsequent evaluation of properties, as we later show for the mechanical properties.

\begin{figure}[!htb]  
  \centering
  \includegraphics[width=0.99\linewidth]{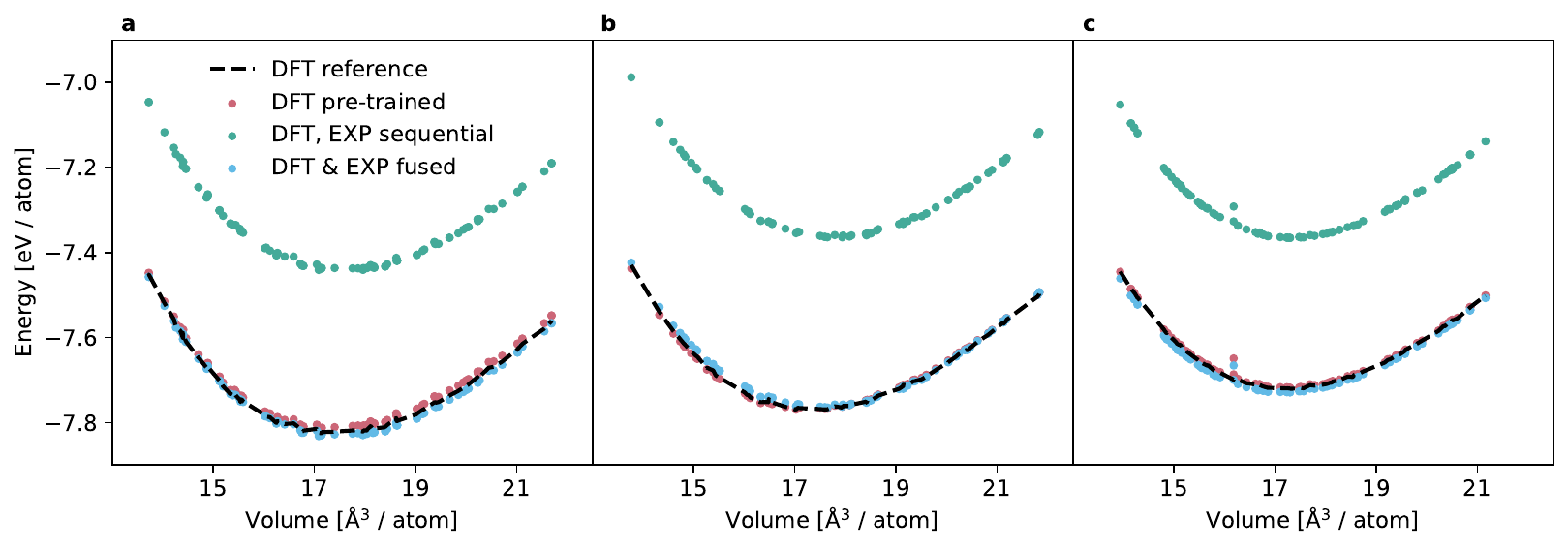} 
   \caption{Energy vs. volume for the hcp (\textbf{a}), fcc (\textbf{b}), and bcc (\textbf{c}) titanium crystal structures computed for samples in the test DFT dataset. The predictions of the DFT pre-trained, DFT, EXP sequential, and DFT \& EXP fused models are denoted with red, green, and blue points, respectively. DFT calculations are denoted with a black dashed line.}
\label{fig:energy_over_volume}
\end{figure}

For the DFT, EXP sequential model, the force and virial errors are still comparable to the DFT pre-trained model but the energy error is drastically increased. This is not surprising considering that MD simulations and our target experimental properties do not depend on energy but only its derivatives. The EXP trainer, therefore, leaves the energy undetermined up to a constant as confirmed by the predicted vs. DFT energy plot (Supplementary Fig. S1). Consequently, any energy-related quantity will also be predicted incorrectly. For example, the energy versus volume equation of state curves for hcp, fcc, and bcc structures are all shifted by a constant and equal value  (Fig.~\ref{fig:energy_over_volume}).  
%%%%%%%%%%%%%%
%%%%%%%%%%%%%%
The DFT, EXP sequential model demonstrates the importance of including DFT data in training, especially when the experimental dataset does not include properties directly related to energies. 

\begin{figure}[!htb]  
  \centering
  \includegraphics[width=0.99\linewidth]{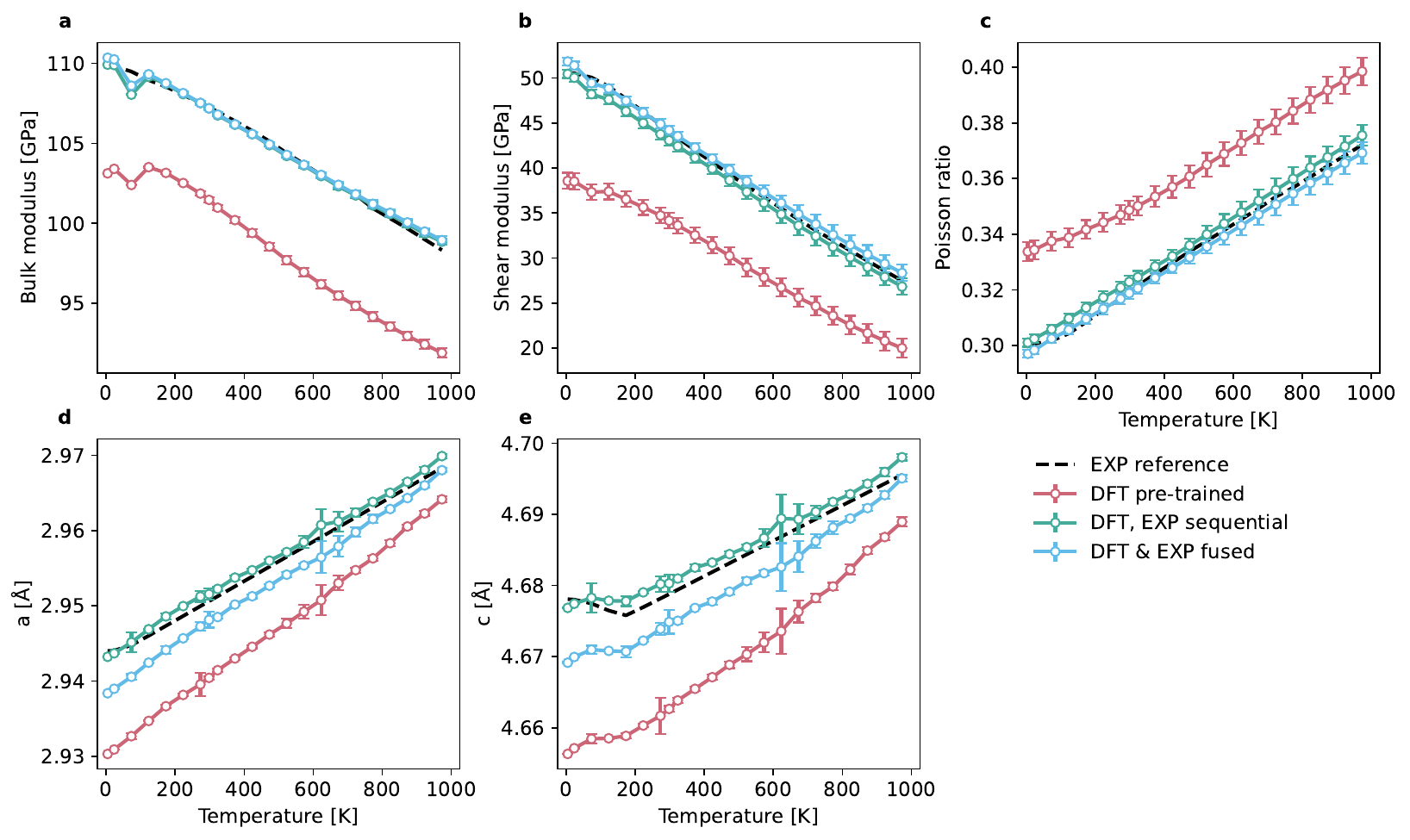} 
   \caption{Bulk modulus (\textbf{a}), shear modulus (\textbf{b}), Poisson's ratio (\textbf{c}), and lattice constants $a$ (\textbf{d}) and $c$ (\textbf{e}) as a function of temperature for hcp titanium. The DFT pre-trained, DFT, EXP sequential, and DFT \& EXP fused models are denoted with red, green, and blue line points, respectively. The experimental results are denoted with a black dashed line. Error bars denote the standard deviation computed via block-averaging with 10 blocks.}
\label{fig:bulk_shear_poisson_lattice}
\end{figure}

%%%%%%%%%%%%%%%%%%%%%%%%%%%%%%%%%%%%%%%%%%%%%%%%%%%%%%%%%%%%%%%%%%%%%%
Next, we evaluate the elastic constants of hcp titanium (Supplementary Fig. S2), which are the target properties of the EXP trainer. Additionally, we report in Fig.~\ref{fig:bulk_shear_poisson_lattice} a-c the bulk modulus, shear modulus, and Poisson's ratio, which are all directly related to elastic constants. 
%%%%%%%%%%%%%%
%%%%%%%%%%%%%%
These properties are computed for all 22 temperatures in the range of 4-973~K where experimental data is available. Training only on DFT data (DFT pre-trained model) fails to reproduce the mechanical properties. On average, the model deviates from the experimental data by 6, 24, and 9$\%$ in bulk modulus, shear modulus, and Poisson's ratio, respectively. In terms of elastic constants, the predictions are for some components off by more than 20 GPa (Supplementary Table S4). Similar deviations in mechanical properties were reported for other ML potentials~\cite{wen2021specialising, dickel2020neural, li2020complex}. Per contra, for the two models that include the EXP trainer, the elastic constants are within a few GPa of the experimental values while the relative errors for the bulk modulus, shear modulus, and Poisson's ratio are below 3$\%$. We obtain a good agreement with experimental observations on the entire investigated temperature range, even though we fit the elastic constants only at four temperatures. Naturally, the agreement is better for the DFT, EXP sequential model because the DFT and experimental datasets are erroneous and, thus, somewhat incompatible. 

An additional target property of the EXP trainer is zero pressure (Supplementary Fig. S3) at fixed, experimentally determined simulation box sizes. In Fig.~\ref{fig:bulk_shear_poisson_lattice} d, e, we show an equivalent result, i.e., the temperature-dependant lattice constants evaluated in the isothermal-isobaric ensemble. Similarly, as for the mechanical properties, the addition of the EXP trainer improves the results for both target and non-target temperatures with DFT, EXP sequential model being the closest to experimental reference values. 

%%%%%%%%%%%%%%%%%%%%%%%%%%%%%%%%%%%%%%%%%%%%%%%%%%%%%%%%%%%%%%%%%%
\subsection*{Generalization to off-target properties and thermodynamic states}
\begin{figure}[!htb]
  \centering
  \includegraphics[width=0.99\linewidth]{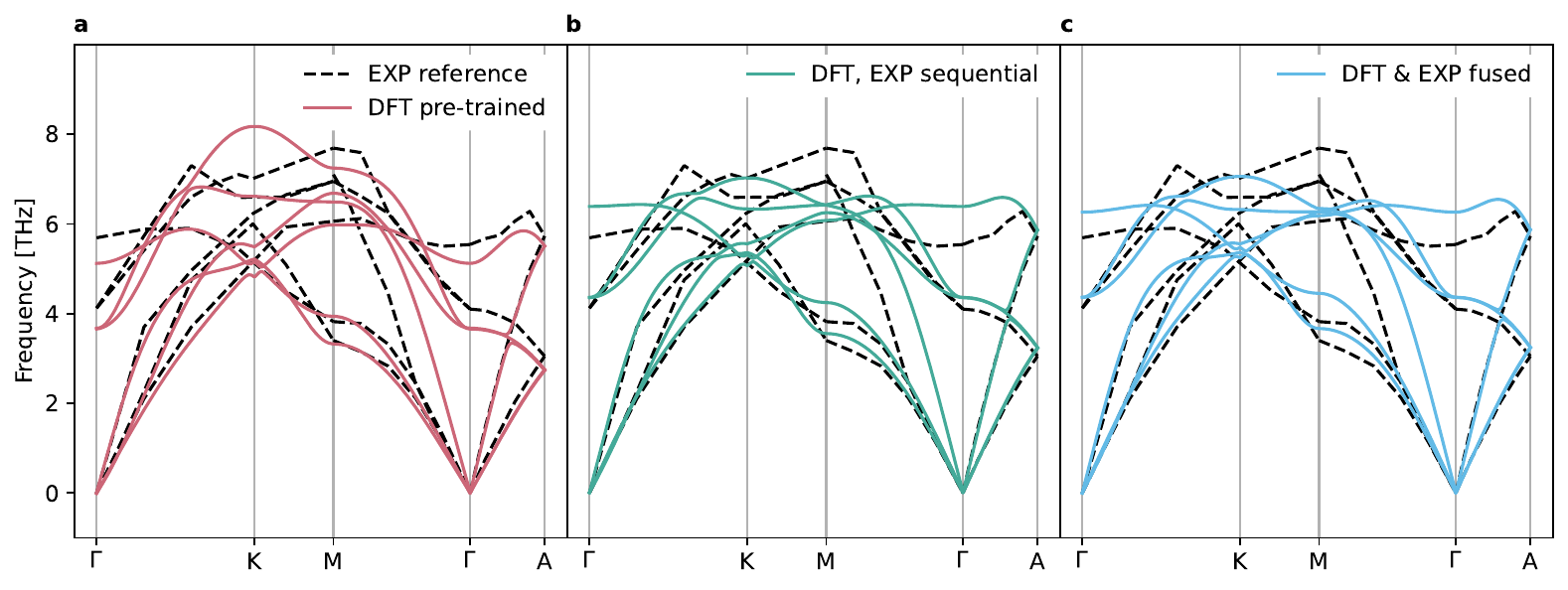} 
\caption{Phonon dispersion curves of hcp titanium for DFT pre-trained (\textbf{a}), DFT, EXP sequential (\textbf{b}), and DFT \& EXP fused (\textbf{c}) models. The ML potential models' predictions match well the black dashed lines denoting the experimental prediction measured at 295~K~\cite{stassis1979lattice}.}
\label{fig:phonon_plots}
\end{figure}

We select the phonon spectra of hcp titanium (Fig.~\ref{fig:phonon_plots}) to test the generalization capabilities to off-target properties. 
%%%%%%%%%%%
%%%%%%%%%%%
All models agree well with experimental prediction. This result is expected for the DFT pre-trained and DFT \& EXP fused models since ML potentials trained on DFT data typically reproduce the phonon dispersion curves well and much better than the classical potentials~\cite{takahashi2017conceptual, bartok2018machine, seko2020machine,wen2021specialising}. Interestingly, we obtain a good agreement also for the DFT, EXP sequential model. Our previous study~\cite{thaler2021learning} showed that training randomly initialized ML potentials on mechanical properties leads to models with drastically different phonon densities of states, i.e., high-capacity models are underconstrained when trained on a small set of target properties. Additional properties could be included to converge toward a unique potential energy solution. However, the required experimental database size is unknown a priori. Our results in Fig.~\ref{fig:phonon_plots} b indicate an alternative route. Pretraining on DFT data seems to constrain the solution to a particular region in parameter space which is only locally modified by the subsequent training on the experimental data. This hypothesis is in accordance with the observed similar force errors for DFT pre-trained and DFT, EXP sequential models (Table~\ref{tab:RMSE_InitAndBulk}).

\begin{figure}[htb!]
    \centering
    \includegraphics[width=0.99\textwidth]{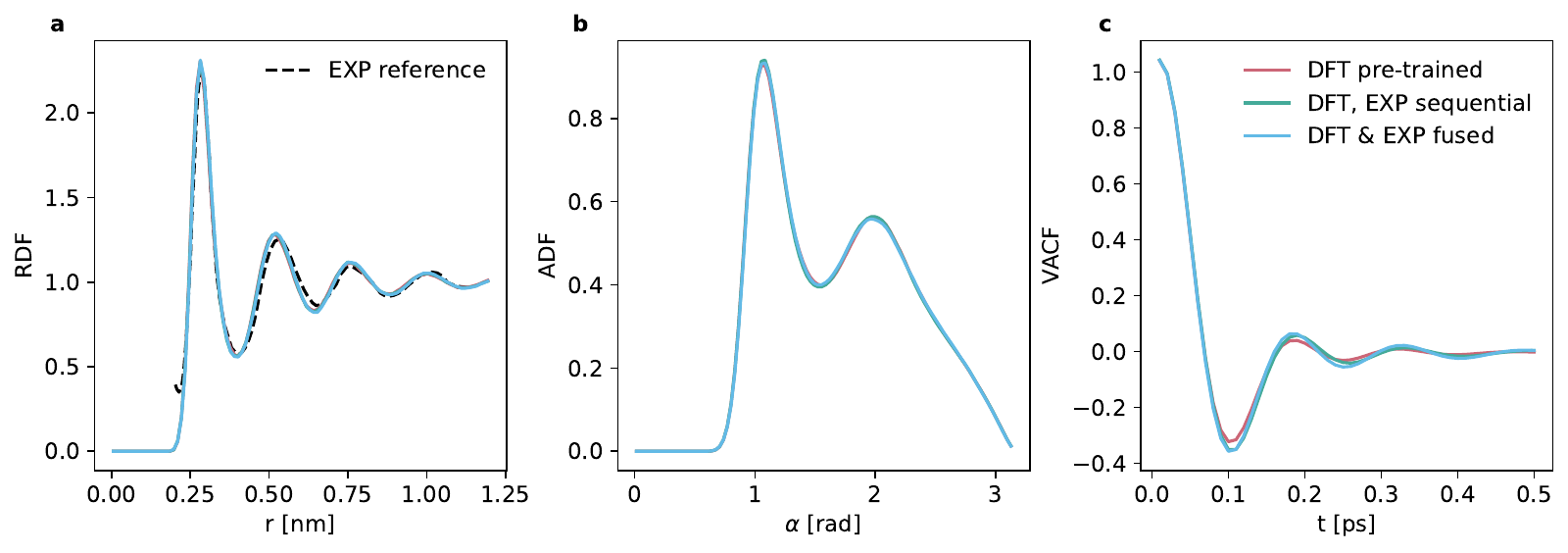}
    \caption{Radial distribution function (RDF, \textbf{a}), angular distribution function (ADF, \textbf{b}), and velocity autocorrelation function (VACF, \textbf{c}) for the DFT pre-trained (red), DFT, EXP sequential (green), and DFT \& EXP fused (blue) models. The RDFs and ADFs are computed at 1965~K to enable direct comparison with experimentally determined RDF (black, dashed)~\cite{holland2007short}. VACFs are evaluated at 2000~K to facilitate comparison to experiments for the self-diffusion coefficient.}
    \label{fig:rdf_adf_vacf}
\end{figure}

To further validate our rationale, we examine the liquid-state titanium's structural and dynamical properties. Two-body and three-body local structural order is measured with radial distribution function (RDF) and angular distribution function (ADF). For all investigated models, the results are indistinguishable within the line thickness (Fig.\ref{fig:rdf_adf_vacf} a, b). 
%%%%%%%%%%%
%%%%%%%%%%%
Moreover, the obtained RDFs are very close to the experimental measurement. Similar conclusions can be drawn for the dynamical properties evaluated via velocity autocorrelation function (VACF), i.e., all models give very similar profiles (Fig.\ref{fig:rdf_adf_vacf} c). We use the VACFs to compute the self-diffusion coefficients and compare them to the experimental value of $(5.3 \pm 0.2) \times 10^{-9}$~m$^2$s$^{-1}$~\cite{meyer2009self}. The ML potentials' results are in good agreement. In particular, we obtain $(5.70 \pm 0.01)$, $(5.19 \pm 0.01)$, and $(5.05 \pm 0.03) \times 10^{-9}$~m$^2$s$^{-1}$ for DFT pre-trained, DFT \& EXP fused, and DFT, EXP sequential model, respectively.

Overall the investigated off-target properties suggest that training on experimental data (subsequently or simultaneously) changes the model with respect to target experimental properties, while the off-target properties are mostly left unaltered. By contrast, with classical potentials~\cite{zhang2021multi} or recent two-body correction to a fixed ML potential~\cite{Sakib2023machine}, re-parametrizing a model to fit a given property usually also modifies other properties. This distinction points to an immensely larger capacity of deep ML potentials compared to classical models.

\subsection*{Experimental data ablation}
\begin{figure}[!htb]  
  \centering
  \includegraphics[width=0.99\linewidth]{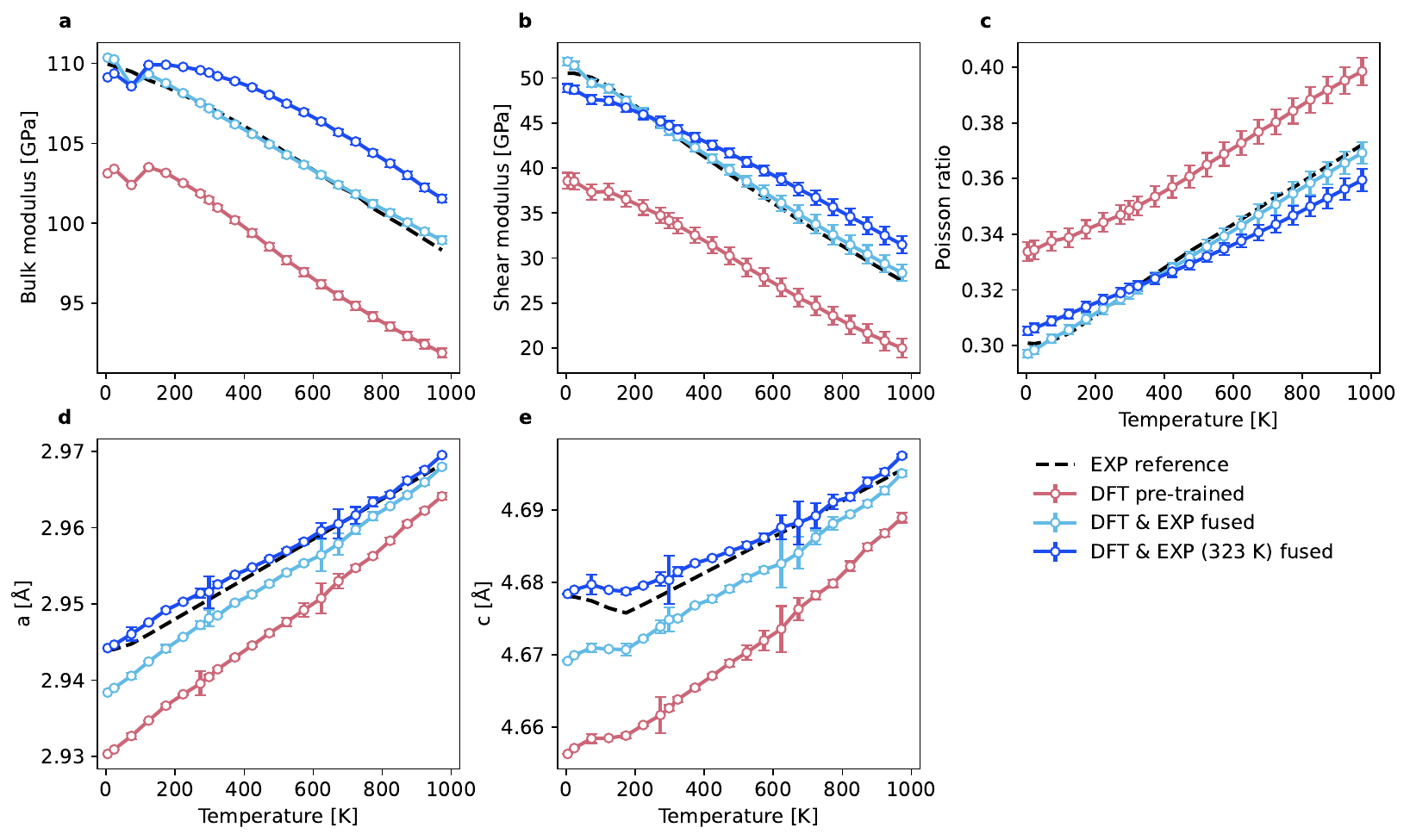} 
   \caption{Bulk modulus (\textbf{a}), shear modulus (\textbf{b}), Poisson's ratio (\textbf{c}), and lattice constants $a$ (\textbf{d}) and $c$ (\textbf{e}) as a function of temperature for hcp titanium. The DFT pre-trained, DFT \& EXP fused, and DFT \& EXP (323~K) fused models are denoted with red, blue, and dark blue line points, respectively. The last two models differ only in experimental training data, i.e., the DFT \& EXP (323~K) fused model is trained on data at a single temperature of 323~K. The experimental reference values are marked with a black dashed line. Error bars denote the standard deviation computed via block-averaging with 10 blocks.}
\label{fig:bulk_shear_poisson_T_transferability}
\end{figure}

Lastly, we consider a data ablation study. An additional model, labeled DFT \& EXP (323~K) fused, is trained with the same approach as the DFT \& EXP fused model, but with experimental training data containing elastic constants and pressure only at a single temperature of 323~K. The aim is to reveal the effect of experimental data size as well as the model's temperature transferability. As shown in Fig.~\ref{fig:bulk_shear_poisson_T_transferability}, the DFT \& EXP (323~K) fused model yields improved mechanical properties and lattice parameters on the entire temperature range compared to training only on DFT data, i.e., DFT pre-trained model. 
%%%%%%%%%%%%%%
%%%%%%%%%%%%%%
 The predicted elastic constants are shown in Supplementary Fig.~4. However, as expected, the mechanical properties are not as accurate as training on experimental data at four different temperatures (DFT \& EXP fused model trained at 23, 323, 623, 923~K). In general, due to the temperature transferability of the models, it seems more beneficial to enlarge the experimental dataset with diverse properties rather than with a single property at densely sampled temperatures. 

%%%%%%%%%%%%%%%%%%%%%%%%%%%%%%%%%%%%%%%%%%%%%%%%%%%%%%% 
\section*{METHODS}
\subsection*{ML potential architecture}
We employ a message passing graph neural network DimeNet++~\cite{gasteiger2020fast} using our implementation in JaxMD~\cite{thaler2021learning}, which takes advantage of neighbor lists for efficient computation of the sparse atomic graph. We select the same neural network hyperparameters (Supplementary Information, Table S1) as in the original publication~\cite{gasteiger2020fast} except for the embedding sizes, which we reduced by factor 4 for computational speed-up. The cut-off is set to 0.5~nm. 

%%%%%%%%%%
\subsection*{DFT trainer}
We use a weighted mean squared error loss function
\begin{equation}
L_{\text{DFT}} = \frac{1}{N_{data}} \sum_{i=1}^{N_{data}} \left[\omega_U (U_i - \tilde{U}_i)^2 +  \frac{\omega_F}{3N_{atoms}}  \sum_{j=1}^{N_{atoms}}\sum_{k=1}^{3}(F_{ijk} - \tilde{F}_{ijk})^2 + \frac{\omega_{V}}{9}\sum_{k=1}^{3}\sum_{l=1}^{3} (V_{ikl} - \tilde{V}_{ikl})^2 \right]
\label{eq:DFTloss}
\end{equation}
where $U_i$ is the energy of the i-th atomic environment in a batch, $F_{ijk}$ is the force in the k-direction of the j-th atom, and $V_{ikl}$ is the virial in the k,l-direction. The reference DFT values are denoted with $\tilde{\;}$. The weights for the energy and force are set to $\omega_U = 1e^{-6}$ and $\omega_F = 1e^{-2}$, while for the virial contribution, only the uniformly deformed supercells contribute with $\omega_V = 4e^{-6}$. The numerical optimization hyperparameters are reported in the Supplementary Information, Table S2.

%%%%%%%%%%
\subsection*{EXP trainer}
We define the loss function as
\begin{equation}
\begin{aligned}
L_{\text{EXP}} &=  \frac{1}{N_{temp}} \sum_{n=1}^{N_{temp}} \sum_{m=1}^{N_{obser}} w_m (O_{m,n} - \tilde{O}_{m,n})^2\\
&=  \frac{1}{N_{temp}} \sum_{n=1}^{N_{temp}} \bigg[ \omega_P (P_n)^2 \\
&+ \frac{\omega_C}{5} \left\{ (C_{11,n} - \tilde{C}_{11,n})^2 + (C_{12,n} - \tilde{C}_{12,n})^2 + (C_{13,n} - \tilde{C}_{13,n})^2  + (C_{33,n} - \tilde{C}_{33,n})^2 + (C_{44,n} - \tilde{C}_{44,n})^2\right\} \bigg] ,
\end{aligned}
\end{equation}
where $O_{m,n}$ is a m-th observable at n-th temperature in a batch and $\tilde{\;}$ denotes the experimental value. The observables are scalar pressure $P_n$ and elastic constants in Voigt notation $C_{**}$. The weights are $\omega_P=1 e^{-9}$ and $\omega_C = 1e^{-10}$. The gradient of the loss with respect to the parameters of the ML potential is obtained with the DiffTRe method~\cite{thaler2021learning}, where the ensemble average of an observable $O_{m,n}$ is computed with the reweighting ansatz for the canonical ensemble~\cite{Zwanzig1954, Norgaard2008, Li2011}
\begin{equation}
    \langle O_{m,n}(U_{\theta}) \rangle \simeq \sum_{i=1}^{N_{traj}} w_i O_{m,n}(S_i, U_{ \theta}) \quad \mathrm{with} \quad 
    w_i =\frac{e^{-\beta (U_{ \theta}(S_i) - U_{\hat{ \theta}}(S_i)) }}{\sum_{j=1}^N e^{-\beta (U_{ \theta}(S_j) - U_{\hat{ \theta}}(S_j)) }}.
\label{eq:weights_definition}
\end{equation}
The summation runs over the trajectory states/atomic environments $S$, $\beta = 1 / (k_B T)$, $k_B$ is the Boltzmann constant, and $T$ is the temperature. $U_{\hat{ \theta}}$ and $U_{\theta}$ denote the reference and perturbed ML potentials. We initialize the forward trajectory generation for every parameter update. Further details can be found in Ref.~\cite{thaler2021learning}. The numerical optimization hyperparameters are reported in the Supplementary Information, Table S2. 

%%%%%%%%%%
\subsection*{ML potential-driven MD simulations}
All MD simulations are performed in JaxMD~\cite{schoenholz2020jax} using a velocity Verlet integrator with a time step of 0.5~fs. The simulated system contains 256 atoms unless otherwise stated. The mass of titanium atoms is set to 47.867~a.u. For NVT simulations during training and to compute the elastic constants and pressure in postprocessing, we use the Langevin thermostat with a friction constant of 4~/ps. For the remaining postprocessing, we run NVT simulations using a Nose-Hoover thermostat and NPT simulations with a Nose-Hoover thermostat and barostat. For the Nose-Hoover chains, we use a chain length of 5, 2 chain steps, and 3 Suzuki-Yoshida steps, and set the thermostat damping parameter to $\tau = 50~\text{fs}$ and the barostat damping parameter to $\tau = 500~\text{fs}$. The pressure is set to 0.

In the EXP trainer, the elastic constants and pressure are computed from 80~ps NVT simulation, where the first 10~ps are disregarded as equilibration, and the state is saved every 0.1~ps. The isothermal elasticity tensor is computed with the stress-fluctuation method~\cite{van2005isothermal,thaler2021learning}. 

To analyze the properties of trained models, we perform the following simulations. For elastic constants and pressure, we perform a 100~ps NVT equilibration run followed by a 1~ns NVT production run. As in training, the box size is set according to the experimental lattice parameters at a given temperature. The elastic constants are saved every 0.1~ps. The bulk modulus and shear modulus are computed from elastic constants (in Voigt notation)~\cite{simmons1971single, li2018elastic, jafari2011pseudopotential} as $K = \nicefrac{2}{9}(C_{11} + C_{12} + 2C_{13} + \nicefrac{1}{2}C_{33})$ and $G = \nicefrac{1}{30}(12C_{44} + 7C_{11} - 5C_{12} + 2C_{33} - 4C_{13})$. The Poisson ratio is computed with $\sigma = (3K - 2G)/(2G + 6K)$. For lattice constants, we perform a 100~ps NPT equilibration followed by a 100~ps NPT production run, where a state is saved every 0.25~ps. For phonon frequency analysis, we generate a $5\times5\times3$ hcp super cell in Avogadro~\cite{hanwell2012avogadro} and employ Phonopy~\cite{togo2023first, togo2023implementation} to compute the phonon densities via finite displacements of $0.01$~Å. To compute the RDF and the ADF, we perform a 100~ps NPT equilibration at 2400~K, 100~ps NPT equilibration at 1965~K, and 80~ps NVT production run at 1965~K, which we sample every 0.1~ps. For these simulations, we double the box size in each dimension, yielding a total of 2048 atoms. The ADF is computed for all triplets within 0.318~nm. For VACF, we perform 100~ps NPT equilibration at 2400~K, 100~ps NPT equilibration at 2000~K, 100~ps NVT equilibration at 2000~K, and 80~ps NVT production run from which we sample every 0.01~ps. The VACF is computed by averaging over 160 different starting points that are 0.5~ps apart. We use the Green-Kurbo relation to compute the self-diffusion~\cite{green1954markoff, kubo1957statistical}. The errors are estimated with block-averaging using 10 blocks.

%%%%%%%%%%%%%%%%%%%%%%%%%%%%%%%%%%%%%%%%%%%%%%%%%%%%%%% 
\subsection*{DATA AVAILABILITY}
The dataset will be made publicly available at \href{https://github.com/tummfm/Fused-EXP-DFT-MLP/tree/main/Dataset}{https://github.com/tummfm/Fused-EXP-DFT-MLP/tree/main/Dataset} upon acceptance of the paper.

\subsection*{CODE AVAILABILITY} 
The code will be made publicly available at \href{https://github.com/tummfm/Fused-EXP-DFT-MLP.git}{https://github.com/tummfm/Fused-EXP-DFT-MLP.git} upon acceptance of the paper.

\bibliography{bibliography}

\subsection*{AUTHOR CONTRIBUTIONS} 
J.Z. conceptualized the study. S.R. implemented and applied the methods and conducted MD simulations as well as postprocessing. S.R. and J.Z. planned the study, analyzed and interpreted results, and wrote the paper.

\subsection*{COMPETING INTERESTS}
The authors declare no competing interests.

\end{document}